\documentclass{elsart1p}
\usepackage{graphicx}
\usepackage{amssymb}
\begin{document}
\begin{frontmatter}
% use the corauthref command within \author for corresponding author
% footnotes;
\title{Neutron Stars as a Probe for Dense Matter}
\author[FIAS]{V.A. Dexheimer}
\author[FIAS,CSC]{S. Schramm}
\address[FIAS]{FIAS, Johann Wolfgang Goethe University, Frankfurt am Main, Germany}
\address[CSC]{CSC, Johann Wolfgang Goethe University, Frankfurt am Main, Germany}
\begin{abstract}
We study different stages of the neutron star cooling by computing neutron star properties at various
temperatures and entropies using an effective chiral model including hadronic and quark degrees of freedom.
Macroscopic properties of the star such as mass and radius are calculated and compared with observations.
It can be seen that the effects of chiral restoration and deconfinement to quark matter in the core of the neutron star at different stages of the evolution can be significant for the evolution of the star and allow insight into the behavior of matter at extreme densities.
\end{abstract}
\end{frontmatter}
%
% main text

Neutron stars that have both hadronic and quark phases are normally described by two different equations of state.
They are connected at the point in which the pressure of the quark phase becomes higher than the pressure of
the hadronic phase. In our approach we have instead a combined model of hadrons and quarks and one equation of state for both phases.
With this, we study important features of this transition, like the restoration of chiral symmetry.
For this purpose we extend the generalized hadronic non-linear chiral model \cite{chiral2,eu}
to include quark degrees of freedom. Although the temperature in late-stage neutron stars is smaller than $1\ $MeV,
in proto-neutron stars, right after the original supernova explosion, it can reach values up to $50\ $MeV.
For this reason we examine the behavior of our model at all possible densities and temperatures.

The lagrangian density of the non-linear sigma model, evaluated in mean field approximation, reads:
\begin{eqnarray}
&L = L_{Kin}+L_{Int}+L_{SB}+L_{Self}-U,&
\end{eqnarray}
where besides the kinetic energy term for baryons, quarks and leptons (included to insure charge neutrality), the terms:
\begin{eqnarray}
&L_{Int}=-\sum_i \bar{\psi_i}[g_{i\omega}\gamma_0\omega+g_{i\phi}\gamma_0\phi+g_{i\rho}\gamma_0\tau_3\rho+m_i^*]\psi_i, \ \ \ \ \ \ \ L_{SB}= m_\pi^2 f_\pi\sigma,&\nonumber
\end{eqnarray}
\begin{eqnarray}
&L_{Self}=-\frac{1}{2}(m_\omega^2\omega^2+m_\rho^2\rho^2+m_\phi^2\phi^2)
-g_4\left(\omega^4+\frac{\phi^4}{4}+3\omega^2\phi^2+\frac{4\omega^3\phi}{\sqrt{2}}+
\frac{2\omega\phi^3}{\sqrt{2}}\right)\nonumber&\\&+\frac{1}{2}k_0(\sigma^2+\zeta^2)-k_1(\sigma^2+\zeta^2)^2
-k_2\left(\frac{\sigma^4}{2}
+\zeta^4\right)
-k_3\sigma^2\zeta
-\epsilon\ \ln{\frac{\sigma^2\zeta}{\sigma_0^2\zeta_0}},&
\end{eqnarray}

represent the interactions between baryons or quarks and vector and scalar mesons, an explicitly chiral symmetry breaking term (responsible for producing the masses of the pseudo-scalar mesons),
the self interactions of scalar and vector mesons and a potential term $U$ for the Polyakov loop as defined below. The index $i$ denotes nucleons or up and down quarks. Note that although we only consider non-strange baryons and quarks in this study, the model includes couplings of the nucleons to the strange scalar condensate field $\zeta$, which is taken into account here. The effective masses of baryons and quarks are generated by the scalar fields except for a small explicit mass term (equal to $\delta m=150$ MeV for baryons and $\delta m=0$ for quarks) and the term containing the Polyakov field $\Phi$:
\begin{eqnarray}
&m_b^*=g_{b\sigma}\sigma+g_{b\zeta}\zeta+\delta m_b+g_{b\Phi} \Phi^2, \ \ \ \  m_q^*=g_{q\sigma}\sigma+g_{q\zeta}\zeta+\delta m_q+g_{q\Phi} (1-\Phi).&
\end{eqnarray}

\begin{figure}
\begin{center}
\includegraphics*[width=6.5cm]{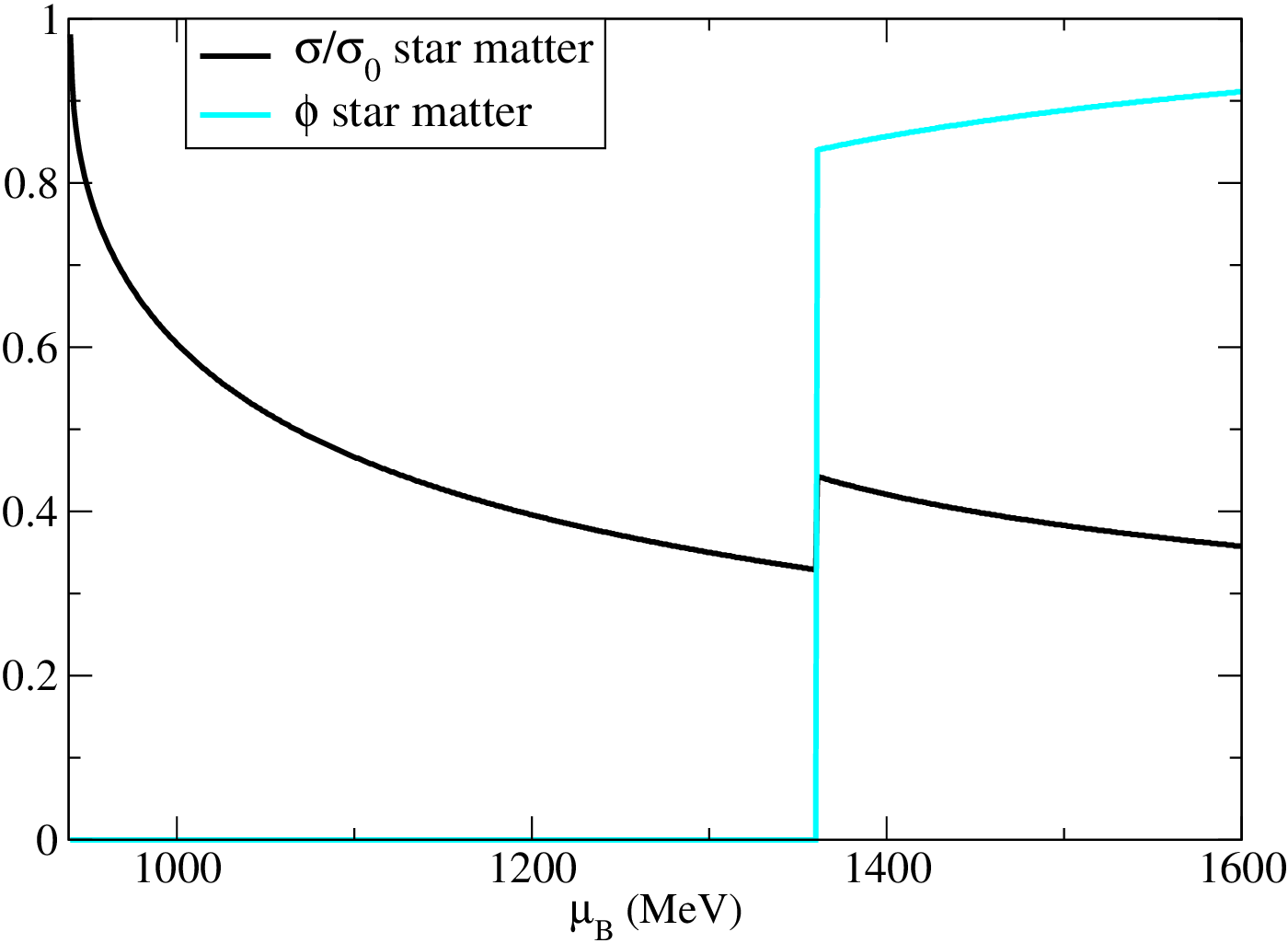}
\includegraphics*[width=6.85cm]{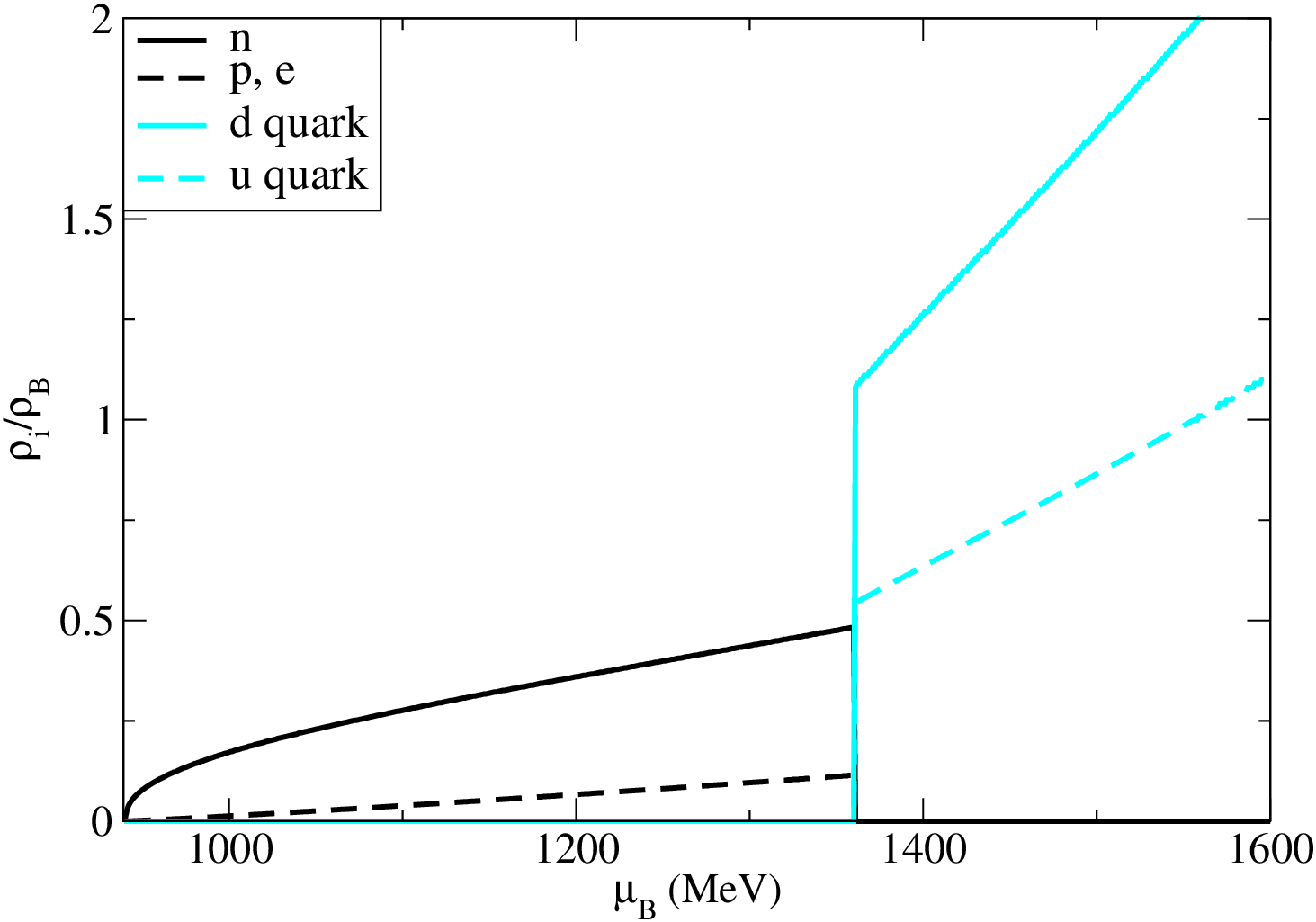}
\end{center}
\caption{Left: Order parameters for chiral symmetry restoration ($\sigma$) and for deconfinement ($\Phi$) for charge neutral and beta equilibrium matter at zero temperature. Right: Population for charge neutral and beta equilibrium matter at zero temperature}
\label{polpop}
\end{figure}

As can be seen from the formula above, there are two order parameters in our model.
The scalar condensate ($\sigma$) decreases with the increase of temperature/chemical potential reducing
the effective masses and signaling the chiral symmetry restoration. The Polyakov loop ($\Phi$)
increases with increasing temperature/chemical potential, which in turn generates higher (lower)
effective masses of the baryons (quarks), effectively describing deconfinement.
This behavior can be seen for neutron star matter in Fig.\ref{polpop}.
The Polyakov loop is related to the Z(3) symmetry, that is spontaneously broken by the presence of quarks.
It has the form $\Phi=\frac13$Tr$[\exp{(i\int d\tau A_4)}]$, where $A_4=iA_0$
is the temporal component of the SU(3) gauge field.

The potential for the Polyakov loop $U$ reads:
\begin{eqnarray}
&U=(a_0T^4+a_1\mu^4+a_2T^2\mu^2)\Phi^2+a_3T_0^4\ln{(1-6\Phi^2+8\Phi^3-3\Phi^4)},&
\end{eqnarray}
that was based on \cite{PNJL,Ratti1,Ratti2} and adapted to include also terms that depend on the chemical potential.
The coupling constants for the hadrons are chosen to reproduce the vacuum masses of the baryons and mesons and the
nuclear saturation properties. The vacuum expectation values of the scalar mesons are constrained by reproducing the
pion and kaon decay constants. The coupling constants for the quarks  are chosen to reproduce lattice data and known information
about the phase diagram (a first order phase transition for pure gauge at $T=T_c$ and a critical point dividing
a crossover area from a first order phase transition line are successfully described in this ansatz \cite{soon}).

\begin{figure}
\begin{center}
\includegraphics*[width=6.6cm]{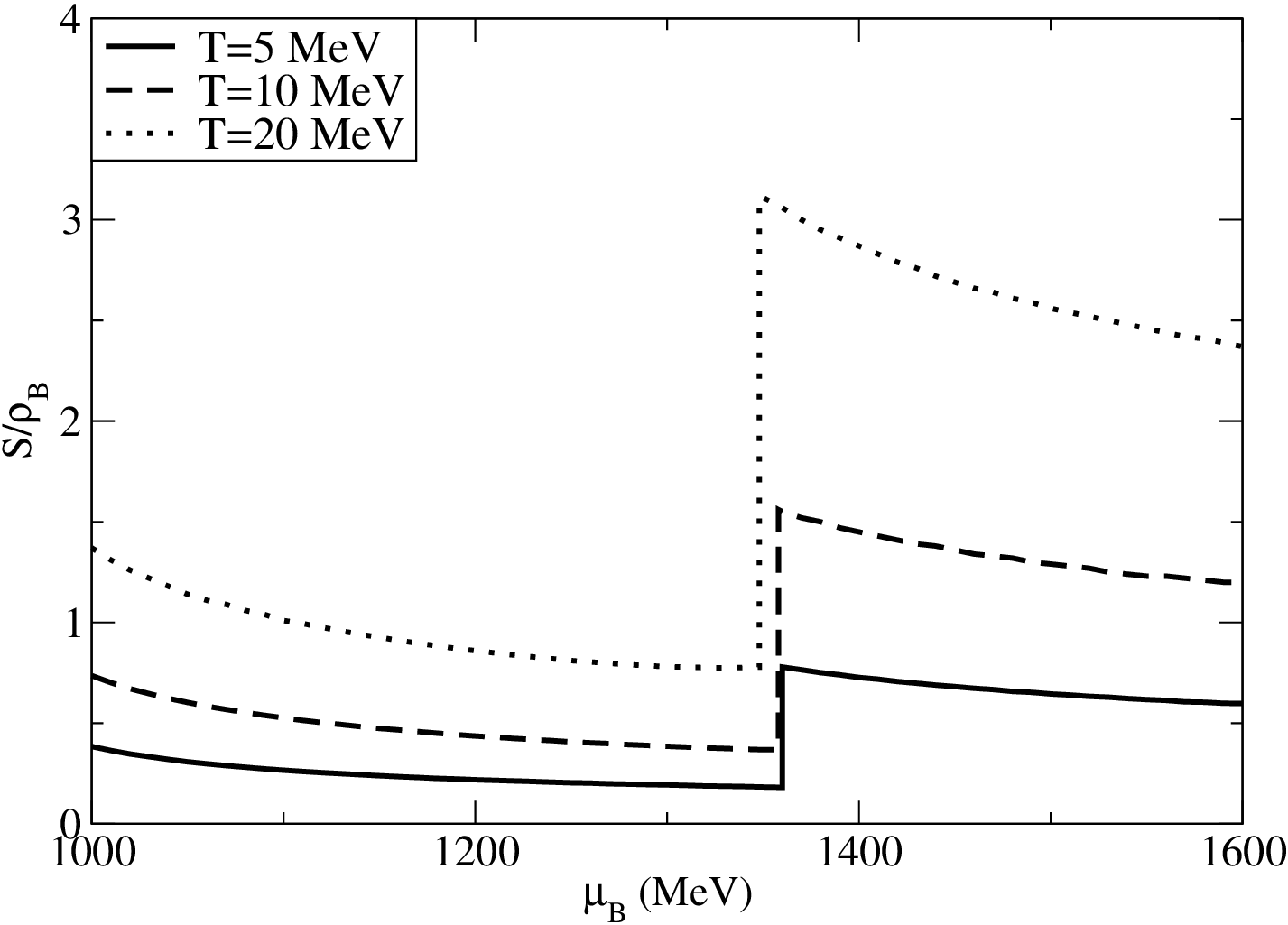}
\includegraphics*[width=6.65cm]{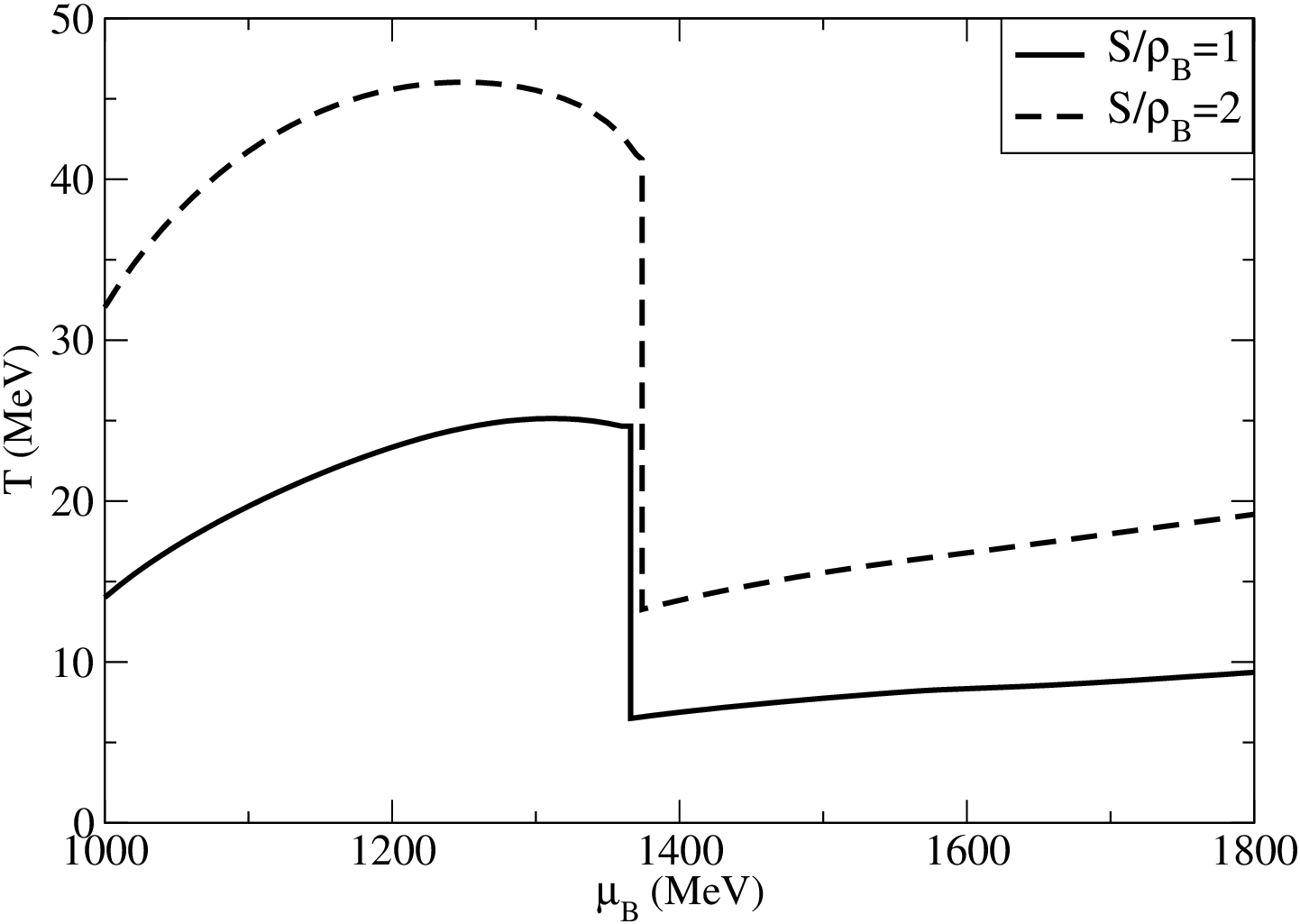}
\end{center}
\caption{Left: Entropy profile for fixed temperatures for charge neutral and beta equilibrium matter. Right: Temperature profile for fixed entropies for charge neutral and beta equilibrium matter}
\label{tempent}
\end{figure}

An interesting effect of the deconfinement is that because of the increase of degrees of freedom,
there is a jump on the entropy at a certain chemical potential for a fixed temperature (Fig. \ref{tempent} left panel).
As a consequence of this, if we fix entropy instead, as normally done in neutron star calculations \cite{eu}, there
will be a jump in the temperature (Fig. \ref{tempent} right panel). This calculations were done requiring charge neutrality independently in each phase.  
If one considers global neutrality instead, the effect of the deconfinement
transition will be smoothed out.

Until now, hyperons were not considered because they appear at about the same densities as quarks and would be
suppressed by the new phase. The strange quark, having a much higher mass, was also not considered because it only appears
at very high densities and would practically not affect neutron star properties. More precise results including both these
features and a full analysis about star stability will be shown in future work \cite{soon}.

\end{document}